\documentclass{jstatphys}
\usepackage[reqno,sumlimits]{amsmath}
\usepackage{amsfonts}
\usepackage{amssymb}
\usepackage{bbm}
\usepackage{psfig}
\usepackage{epsfig}
\def\rz{{\mathbbm{R}}}
\def\gz{{\mathbbm{Z}}}

\begin{document}

\title[Phase transition of the Ising model with fixed magnetization]{Existence and order of the phase transition of the Ising model with fixed magnetization}

\author{Michael Kastner%
\footnote{Institut f{\"u}r Theoretische Physik, Friedrich-Alexander-Universit{\"a}t Erlangen-N{\"u}rnberg, Staudtstra{\ss}e 7, 91058 Erlangen, Germany;\\
New address: INFM Unit\'a di Firenze, Via G.\ Sansone 1, 50019 Sesto Fiorentino (FI), Italy; e-mail: kastner@fi.infn.it}
}
\runningauthor{Michael Kastner}

\date{Version of \today}

\begin{abstract}
Properties of the two dimensional Ising model with fixed magnetization are deduced from known exact results on the two dimensional Ising model. The existence of a continuous phase transition is shown for arbitrary values of the fixed magnetization when crossing the boundary of the coexistence region. Modifications of this result for systems of spatial dimension greater than two are discussed.
\end{abstract}


\section{Introduction}
In a recent paper,\cite{PleiHue} $2d$ and $3d$ Ising models with fixed magnetization are investigated numerically and signatures are found in the microcanonical caloric curves of finite systems which hint a first-order phase transition. From the numerical data, however, it is not clear if the observed signatures scale appropriately to persist in the thermodynamic limit, and hence properties of the infinite system cannot be inferred from the data.

 Although an exact solution is known only for the zero-field case of the $2d$ Ising model {\em without} any constraints on the magnetization,\cite{Onsager} there are a number of exact results which can be used to tackle the questions of existence and order of a phase transition at fixed magnetization analytically. As a convenient thermodynamic function to discuss this topic, the entropy as a function of the interaction energy and the magnetization is chosen.

An important ingredient in the following discussion is the equivalence of ensembles, which holds for Ising systems of arbitrary spatial dimension.\cite{LePfiSul} This allows to combine the known exact results, which are typically obtained in the canonical ensemble, with some simple geometrical arguments on the microcanonical entropy.

Sec.~\ref{ising} gives a definition of the Ising model. In Sec.~\ref{entropy}, several thermodynamic quantities are defined and implications of some known exact results of the $2d$ Ising model on these quantities are discussed. Then, only elementary analysis is needed to establish the existence of a phase transition of the Ising model with fixed magnetization in Sec.~\ref{existence} and to identify the transition as a continuous one in Sec.~\ref{order}. Modifications of this results for the case of spatial dimension $d>2$ are discussed in Sec.~\ref{highdim}.

\section{The $2d$ Ising model}
\label{ising}
\noindent Consider an even%
\footnote{This is only for notational simplicity.}
number $N=L^2$, $L\in2\gz$, of classical spins $\sigma_i\in\{-1,+1\}$, $i=1,\ldots,N$, on a two dimensional quadratic lattice $\{1,\ldots,L\}^2$ with periodic boundary conditions. Then the nearest-neighbor Ising model\cite{Ising} is defined by the Hamiltonian
\begin{equation}
{\mathcal{H}}:\Gamma_N\to\rz,\qquad \sigma\mapsto {\mathcal{E}}(\sigma)-h{\mathcal{M}}(\sigma)
\end{equation}
where $\Gamma_N\equiv\{-1,+1\}^N$ is the configuration space of the system, $\sigma\equiv(\sigma_1,\ldots,\sigma_N)\in\Gamma_N$ are called configurations, and $h\in\rz$ is an external magnetic field.
\begin{equation}
{\mathcal{M}}:\Gamma_N\to(2\gz)\cap[-N,+N],\qquad \sigma\mapsto \sum_{i=1}^N\sigma_i
\end{equation}
is the magnetization and
\begin{equation}
{\mathcal{E}}:\Gamma_N\to(4\gz\cap[-2N,+2N])\setminus\{-2N+8,2N-8\},\qquad \sigma\mapsto \sum_{\langle i,j\rangle} \sigma_i\sigma_j
\end{equation}
is the interaction energy. $\langle i,j\rangle$ denotes a summation over all pairs of spins which are neighbors on the lattice.

\section{Thermodynamic quantities}
\label{entropy}
\noindent We follow Ref.~\cite{Pfister} to define the entropy density%
\footnote{The term ``density'' will be omitted in the following}
$s$ of the $2d$ Ising model in the thermodynamic limit $N\to\infty$ as a function of the interaction energy density\footnotemark[2] $\varepsilon$ and the magnetization density\footnotemark[2] $m$. Let
\begin{equation}
B_\varrho(\varepsilon,m):=\left\{(\varepsilon',m')\in\rz^2:\,(\varepsilon-\varepsilon')^2+(m-m')^2\leq\varrho^2\right\}
\end{equation}
be the ball of center $(\varepsilon,m)$ and radius $\varrho$. Then the entropy can be defined as
\begin{eqnarray}
s&:&{\mathcal D}(s)\to\rz,\\
&&(\varepsilon,m)\mapsto\min_{\varrho>0}\lim_{N\to\infty}\frac{1}{N}\ln\left|\left\{\sigma\in\Gamma_N: \left(\frac{\mathcal{E}(\sigma)}{N},\frac{\mathcal{M}(\sigma)}{N}\right)\in B_\varrho(\varepsilon,m)\right\}\right|\nonumber
\end{eqnarray}
where $|\cdot|$ denotes the cardinality of a set. The domain
\begin{equation}
{\mathcal D}(s):=\left\{(\varepsilon,m)\Big|\,\varepsilon\in[-2,+2],\, |m|\leq\frac{2-\varepsilon}{4}\right\}
\end{equation}
of the entropy has the shape of a triangle. This is due to the fact that a given value of the interaction energy $\varepsilon$ is equivalent to a given proportion of ``antiparallel'' neighboring spins $\sigma_i=-\sigma_j$ ($i,j$ neighbors), which implies an upper bound on the absolute value of the magnetization for configurations of the given energy. Due to the spin inversion symmetry of the Ising Hamiltonian, the entropy $s(\varepsilon,m)$ is symmetric with respect to the magnetization $m$.%
\setlength{\floatsep}{-12cm}
\begin{figure}[h]
\vspace{-4.5cm}
\centerline{
\hspace{0.5cm}
\psfig{figure=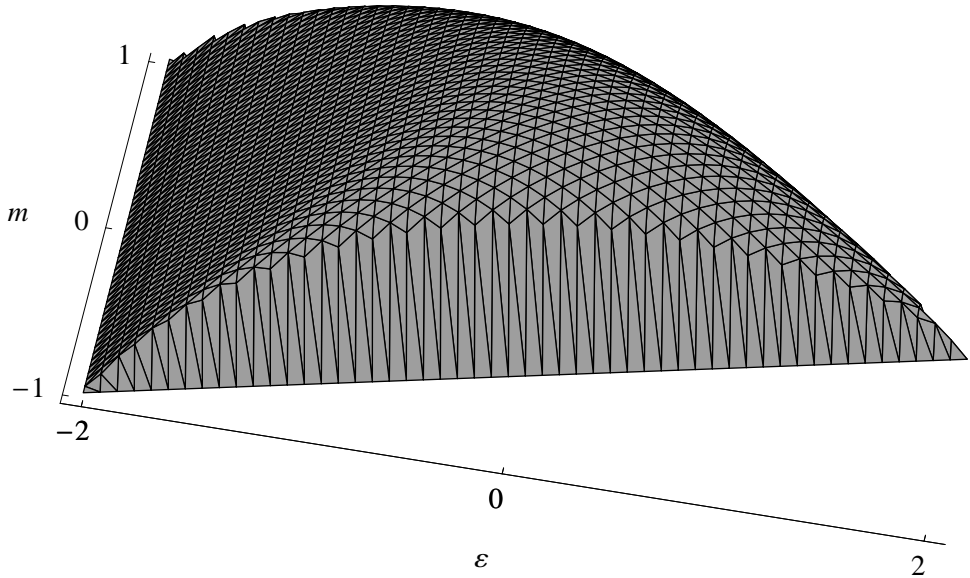,width=19cm}}
\vspace{-14.4cm}\caption{\label{ent2d50} \small Approximate graph of the entropy as a function of the interaction energy $\varepsilon$ and the magnetization $m$.}
\end{figure}
Fig.~\ref{ent2d50} is intended to give an idea how the graph of $s$ approximately looks like. An exact closed form expression for $s$ is not known. Via Legendre transformation, this would be equivalent to a closed form expression of the free energy density of the $2d$ Ising model for arbitrary external field $h$.

The variable thermodynamically conjugate to the magnetization $m$ is $\beta h$, the product of the inverse temperature $\beta$ and the external field $h$. An expression of the magnetization
\begin{equation}
\widetilde{m}:[-2,+2]\times(\rz\setminus\{0\})\to[-1,+1],\qquad(\varepsilon,\beta h)\mapsto\widetilde{m}(\varepsilon,\beta h)
\end{equation}
as a function of $\beta h$ and the interaction energy $\varepsilon$ is obtained from the entropy implicitly via the Legendre transformation
\begin{equation}\label{mvonh}
\sup_{m}[s(\varepsilon,m)+\beta h m]=:s(\varepsilon,\widetilde{m}(\varepsilon,\beta h))+\beta h \widetilde{m}(\varepsilon,\beta h)
\end{equation}
The spontaneous magnetization $m_\pm$ is defined%
\footnote{This is not entirely in accordance with the standard terminology, where one speaks of spontaneous magnetization only where $m_\pm(\varepsilon)\neq0$.}
as the zero-field limit
\begin{equation}\label{m_spont}
m_\pm:[-2,+2]\to[-1,+1],\qquad\varepsilon\mapsto\lim_{\beta h\to\pm 0}\widetilde{m}(\varepsilon,\beta h)
\end{equation}
\begin{figure}[ht]
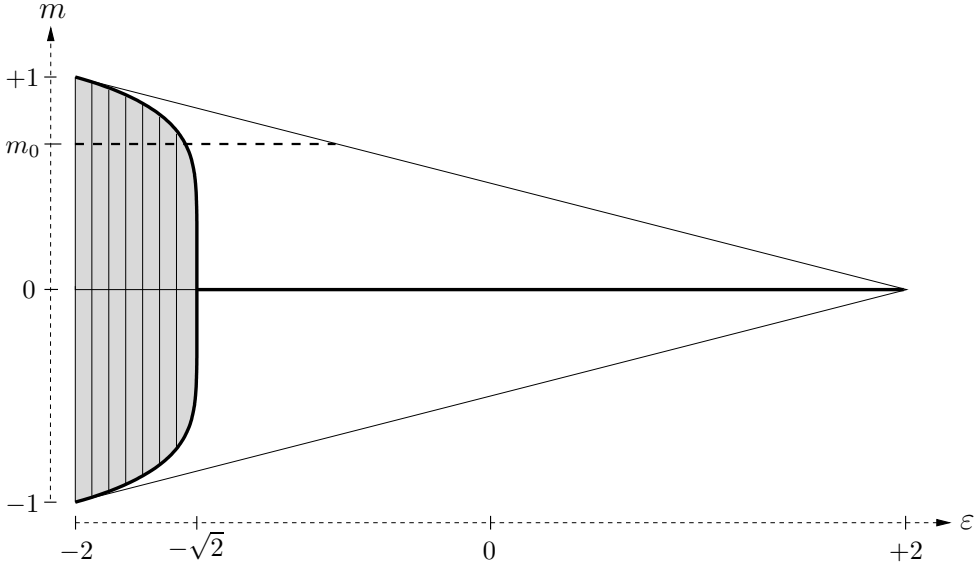

\include{contour}
\caption{\label{contour} \small Schematic picture of the entropy $s(\varepsilon,m)$. The triangle is the domain of the entropy, the bold line is the spontaneous magnetization. Parallels drawn inside the coexistence region $C$ (hatched) indicate straight lines of constant entropy.}
\end{figure}
The domain of the entropy $s$, the triangle ${\mathcal D}(s)$, will be used in the schematic picture of Fig.~\ref{contour} to illustrate some properties of the entropy. The bold curve in the triangle is the spontaneous magnetization of the $2d$ Ising model, for which an exact expression is known from combining Yang's result\cite{Yang} with the caloric curve obtained from Onsager's solution.\cite{Onsager} For the spontaneous magnetization the condition 
\begin{equation}\label{msp=0}
m_\pm(\varepsilon)=0\Leftrightarrow\varepsilon\geq-\sqrt{2}
\end{equation}
holds, and the value $-\sqrt{2}$ is called the critical energy.

The region of ${\mathcal D}(s)$ bounded by the non-zero spontaneous magnetization and the line $\varepsilon=-2$,
\begin{equation}
C:=\Big\{(\varepsilon,m)\in {\mathcal D}(s)\,\Big|\,|m|<|m_\pm(\varepsilon)|\Big\}
\end{equation}
is called the coexistence region (hatched in Fig.~\ref{contour}). Inside the closure $\overline{C}$ of the coexistence region, $s$ is constant with respect to the magnetization, i.e.,
\begin{equation}\label{constant}
s(\varepsilon,m)=s(\varepsilon,m_\pm(\varepsilon))\quad\forall(\varepsilon,m)\in\overline{C}
\end{equation}
which is the signature of the field-driven first-order phase transition of the $2d$ Ising model at low energies and zero external field. In Fig.~\ref{contour}, some parallel lines are drawn inside the coexistence region to symbolize lines of constant entropy. Outside the coexistence region $C$, the concavity of the entropy combined with the definition of the spontaneous magnetization (\ref{m_spont}) implies the strict inequality
\begin{equation}\label{less}
s(\varepsilon,m)<s(\varepsilon,m_\pm(\varepsilon))\quad\forall(\varepsilon,m)\notin\overline{C},\,m\neq0
\end{equation}
Now, from the two-variable function $s$, we define the one-variable function
\begin{equation}
s_{\mathrm{zf}}:{\mathcal D}(s_{\mathrm{zf}})\to\rz,\qquad\varepsilon\mapsto s(\varepsilon,m)\Big|_{m=m_\pm(\varepsilon)}
\end{equation}
on the domain ${\mathcal D}(s_{\mathrm{zf}})=[-2,+2]$, which is the entropy of the $2d$ Ising model in the \underline{z}ero-\underline{f}ield limit, and the family of one-variable functions
\begin{equation}
s_{m_0}:{\mathcal D}(s_{m_0})\to\rz,\qquad\varepsilon\mapsto s(\varepsilon,m)\Big|_{m=m_0}
\end{equation}
on the domain ${\mathcal D}(s_{m_0})=[-2,2-4|m_0|]$, which is the entropy of the $2d$ Ising model with fixed magnetization $m_0$. In the following, existence and order of a phase transition in the $2d$ Ising model with fixed magnetization are discussed by making use of these two functions.

\section{Existence of a phase transition}
\label{existence}
\noindent For all values of the fixed magnetization $|m_0|<1$, the entropy $s_{m_0}$ is non-analytic for at least one value of $\varepsilon\in{\mathcal D}(s_{m_0})$. This is a consequence of:
\renewcommand{\labelenumi}{\alph{enumi})}
\begin{enumerate}
\item The fact that every line $\left\{(\varepsilon,m)\in \mathrm{int}\,{\mathcal D}(s)\,\big|\,m=m_0\right\}$ of constant magnetization $m_0\in(-1,+1)$ in the interior of ${\mathcal D}(s)$ has at least one point of intersection with the spontaneous magnetization (as sketched in Fig.~\ref{contour}),
\item $\displaystyle s_{m_0}(\varepsilon)=s_{\mathrm{zf}}(\varepsilon)$ for all $(\varepsilon,m_0)\in \overline{C}$ [from Eq.~(\ref{constant})],
\item $\displaystyle s_{m_0}(\varepsilon)<s_{\mathrm{zf}}(\varepsilon)$ for all $(\varepsilon,m_0)\notin \overline{C}$, $m_0\neq0$ [from Eq.~(\ref{less})],
\item The entropy $s_{\mathrm{zf}}(\varepsilon)$ of the Ising model in zero field is analytic for all $\varepsilon<-\sqrt{2}$ and non-analytic for $\varepsilon=-\sqrt{2}$ [Otherwise $\beta(\varepsilon)=\frac{\partial s_{\mathrm{zf}}(\varepsilon)}{\partial \varepsilon}$ would show a non-analyticity for $\varepsilon<-\sqrt{2}$. Due to the equivalence of ensembles, this would be in contradiction to the analytic result for the caloric curve $\varepsilon(\beta)$ below the critical temperature obtained from Onsager's solution.\cite{Onsager}],
\item $\displaystyle s_{m_0=0}(\varepsilon)=s_{\mathrm{zf}}(\varepsilon)$ for all $\varepsilon$ [from Eqs.~(\ref{msp=0}) and (\ref{constant})].
\end{enumerate}

{\bf Case $\mathbf{m_0\neq0}$:} Item (a) guarantees that, for all values of the fixed magnetization $|m_0|<1$, the line of constant $m_0$ crosses the boundary of the coexistence region once, and the corresponding energy $\varepsilon_0(m_0)$ is given by the inverse function of the spontaneous magnetization $m_\pm$. From conditions (b) and (c), we know that $s_{ \mathrm{zf}}$ and $s_{m_0}$ are identical on the interval $[-2,\varepsilon_0(m_0)]$, but different elsewhere. According to (d), $s_{ \mathrm{zf}}$ is analytic on the interval $(-2,-\sqrt{2})$, and, as the analytic continuation of a function is unique, it follows that $s_{m_0}(\varepsilon)$ is non-analytic at $\varepsilon=\varepsilon_0(m_0)< -\sqrt{2}$ for all $0<|m_0|<1$.

{\bf Case $\mathbf{m_0=0}$:} In this case, according to item (e), the entropy function of the Ising model with fixed magnetization is identical to that of the zero field case on the entire interval $[-2,+2]$. Hence, also the non-analyticity of $s_{ \mathrm{zf}}(\varepsilon)$ at $\varepsilon=\varepsilon_0(0)=-\sqrt{2}$ [item (d)] is present in $s_{m_0=0}$.

From Yang's result\cite{Yang} for the spontaneous magnetization of the $2d$ Ising model, the inverse temperature $\beta_0$ corresponding to $\varepsilon_0(m_0)$ via the caloric curve is known to be
\begin{equation}
\beta_0:[-1,+1]\to[\beta_c,\infty],\qquad m_0\mapsto\frac{1}{2}\,\mbox{arcsinh}\left[\left(1-m_0^8\right)^{-1/4}\right]
\end{equation}
where $\beta_c=\frac{1}{2}\ln(1+\sqrt{2})$ is the critical temperature of the $2d$ Ising model. Hence, we conclude that the $2d$ Ising model with fixed magnetization $m_0$ shows a phase transition at the inverse temperature $\beta_0(m_0)$.

\section{Order of the phase transition}
\label{order}
\noindent In this section, we show that the phase transition of the Ising model with fixed magnetization is a continuous one.

A temperature-driven first-order phase transition is characterized by the appearance of latent heat, i.e., a discontinuity in the caloric curve $\varepsilon(\beta)$, where $\beta$ is the inverse temperature. In the Ising model with fixed magnetization, due to the thermal equation of state
\begin{equation}\label{tes}
\beta(\varepsilon,m)=\frac{\partial s(\varepsilon,m)}{\partial \varepsilon}
\end{equation}
such a discontinuity corresponds to a linear piece in the entropy $s_{m_0}$. In the following, by proving the absence of a linear piece in $s_{m_0}$, the phase transition of the $2d$ Ising model with fixed magnetization $m_0$ is shown to be continuous for all values of $m_0\in[-1,+1]$. For this discussion we distinguish between the two cases of a linear piece {\em inside} and {\em outside} the coexistence region, respectively.
\subsection{No linear piece of $s_{m_0}$ inside the coexistence region}\label{nolinin}
\noindent From Eq.~(\ref{constant}) it follows that a linear piece of $s_{m_0}$ inside the coexistence region implies a linear piece in $s_{\mathrm{zf}}$. It is known from Onsager's solution\cite{Onsager} that there is no temperature-driven first-order phase transition in the $2d$ Ising model in zero field, hence there is no linear piece of $s_{m_0}$ inside the coexistence region.
\subsection{No linear piece of $s_{m_0}$ outside the coexistence region}\label{nolinout}
\noindent This result can already be found in the literature. For an Ising system of arbitrary spatial dimension in non-zero external field or at temperatures above the critical temperature, there exists a unique, translation invariant equilibrium state.\cite{LebMar} It is shown in Ref.~\cite{Pfister} that this implies strict concavity of the entropy outside the phase coexistence region $C$. This strict concavity of course does not allow for a linear piece in the entropy $s_{m_0}$ of an Ising system with fixed magnetization $m_0$ outside $C$.

Additionally, another proof of the absence of a linear piece in $s_{m_0}$ outside the coexistence region is given by showing a stronger result on the analyticity properties of the entropy $s$. It will be argued that a linear piece of $s_{m_0}$ outside the coexistence region is in contradiction to the analyticity properties of the Gibbs free energy
\begin{equation}
g:{\mathcal D}(g)\to\rz,\qquad(\beta,\beta h)\mapsto-\lim_{N\to\infty}(\beta N)^{-1}\ln\sum_{\sigma\in\Gamma_N}\mathrm{e}^{-\beta{\mathcal H}(\sigma)}
\end{equation}
${\mathcal D}(g):=\rz^+\!\times\rz$, of the $2d$ Ising model. It follows from the circle theorem of Lee and Yang\cite{LeeYang} and a result from Lebowitz and Penrose\cite{LebPen} on analyticity properties, that $g$ is analytic for all non-zero values of the external field $h$ or for all inverse temperatures $\beta$ smaller than the critical inverse temperature.%
\footnote{Actually, this result is valid for arbitrary spatial dimension.}
The Gibbs free energy $g$ and the entropy $s$ are connected by means of a Legendre transformation
\begin{equation}\label{leg_trafo}
s(\varepsilon,m)=\sup_{\beta,\beta h}\left[-\beta g(\beta,\beta h)+\beta\varepsilon-\beta h m\right]
\end{equation}
For regions where $g$ is differentiable, $\varepsilon$ and $m$ can be expressed in terms of the Gibbs free energy as
\begin{equation}
m(\beta,\beta h)=-\beta\frac{\partial g(\beta,\beta h)}{\partial(\beta h)}
\end{equation}
and
\begin{equation}
\varepsilon(\beta,\beta h)=g(\beta,\beta h)+\beta\frac{\partial g(\beta,\beta h)}{\partial \beta}
\end{equation}
For regions where $g$ is strictly concave, the inversions $\beta(\varepsilon,m)$ and $\beta h(\varepsilon,m)$ exist, and Eq.~(\ref{leg_trafo}) can be rewritten as
\begin{eqnarray}
s(\varepsilon,m)=\beta(\varepsilon,m)\left[\beta(\varepsilon,m)\frac{\partial g(\beta(\varepsilon,m),\beta h(\varepsilon,m))}{\partial\beta(\varepsilon,m)}\right.\nonumber\\ \left.+\beta h(\varepsilon,m)\frac{\partial g(\beta(\varepsilon,m),\beta h(\varepsilon,m))}{\partial\beta h(\varepsilon,m)}\right]
\end{eqnarray}
This implies analyticity of $s(\varepsilon,m)$ for all pairs $(\varepsilon(\beta,\beta h),m(\beta,\beta h))\in P$ for which $g(\beta,\beta h)$ is analytic, and 
\begin{equation}
P=\left\{(\varepsilon,m)\in{\mathcal D}(s)\setminus\overline{C}\Big|\,\frac{\partial s(\varepsilon,m)}{\partial\varepsilon}>0\right\}
\end{equation}
is the set of pairs $(\varepsilon,m)$ outside the closure of the coexistence region corresponding to positive temperatures. Then, also $s_{m_0}(\varepsilon)$ is analytic for all pairs $(\varepsilon,m_0)\in P$.

The entropy $s$ is zero on the boundary of its domain and continuously differentiable on its entire domain\cite{LiebYng}. Together with the analyticity properties discussed above, this rules out the possibility of a linear piece in $s_{m_0}$ outside the coexistence region.

\section{Spatial dimension greater than two}
\label{highdim}
\noindent Some remarks are in order concerning the validity of the arguments of Secs.~\ref{existence} and \ref{order} for Ising models of spatial dimension $d>2$. The crucial difference to the $2d$ Ising model is that, to the knowledge of the author, it is not clear if there exists more than one temperature-driven phase transition in the Ising model in zero field for $d>2$. If this is not the case, the results of this paper are valid for Ising systems of higher spatial dimension. If there is more than one such transition, some modifications of the results on the Ising model with fixed magnetization arise:
\begin{itemize}
\item For values of $(\varepsilon,m_0)$ inside the coexistence region $C$, the Ising model with fixed magnetization shows identical thermal behavior (in the sense of identical caloric curves) as the Ising model in zero field, including phase transitions possibly existing in the zero-field case.
\item The results used in Sec.~\ref{nolinout} are valid for Ising systems of arbitrary spatial dimension. Hence, there is no linear piece in the entropy of the Ising model with fixed magnetization outside the coexistence region $C$.
\item In general the arguments of Sec.~\ref{existence} are also valid for $d>2$ and the Ising model with fixed magnetization shows a phase transition when crossing the boundary of $C$. However, when crossing this boundary at an energy for which the zero-field entropy $s_{\mathrm{zf}}$ is non-analytic, it might be the case that the entropy of the Ising model with fixed magnetization is the analytic continuation of $s_{\mathrm{zf}}$ across the locus of the non-analyticity. Then, for this particular value of the fixed magnetization $m_0$, the Ising model with fixed magnetization does not show a phase transition when crossing the boundary of the coexistence region.
\end{itemize} 

\section{Conclusion}
\noindent In the $2d$ Ising model with fixed magnetization, for all values of the fixed magnetization $|m_0|<1$, there is a continuous phase transition. This phase transition is unique in the sense that there are no further non-analyticities in any thermodynamic function or in the caloric curve. The inverse temperature at which the phase transition occurs is given by $\beta_0(m_0)=\frac{1}{2}\,\mbox{arcsinh}\left[\left(1-m_0^8\right)^{-1/4}\right]$. Hence, in the thermodynamic limit, the S-shape reported for the microcanonical caloric curve of finite Ising systems with fixed magnetizations in Ref.~1, does {\em not} converge to a straight line signalling a first order phase transition, but to a single point at which a continuous phase transition occurs.

For spatial dimensions $d>2$, a first-order transition can occur in the Ising model with fixed magnetization only if there is a first order transition in the $d$-dimensional zero-field case. Although, to the knowledge of the author, a proof is lacking, this is expected to be not the case for the $3d$ Ising model.

\section*{Acknowledgements}
\noindent The author would like to thank Alfred H{\"u}ller for stimulating discussions, and Michael Promberger and Simone Warzel for helpful comments on the manuscript. Some references cited were pointed out by Charles-Edouard Pfister and unknown referees. The data of Fig.~\ref{ent2d50} were kindly supplied by Michael Waasner.


\end{document}